\documentclass[12pt]{iopart}
\usepackage{iopams}  
\usepackage{graphicx}
\usepackage{epstopdf}
\usepackage{color}
\newcommand{\CsCuCl}[0]{Cs$_2$CuCl$_4$ }
\newcommand{\CsE}[0]{Cs$_2$CuCl$_4$}

\begin{document}

\setcounter{figure}{0}

\title[Low energy excitations in \CsCuCl]
{Nature of the low energy excitations in the short range ordered region of \CsCuCl as revealed by $^{133}$Cs NMR}

\author{M. -A. Vachon, G. Koutroulakis, V. F. Mitrovi{\'c}, Ookie Ma, and J .B. Marston}
\address{Department of Physics, Brown University, Providence, RI 02912, U.S.A. }

 \author{A. P. Reyes and P. Kuhns}
  \address{National High Magnetic Field Laboratory, Tallahassee, FL 32310, U.S.A.}
  \author{R. Coldea}
\address{Clarendon Laboratory, University of Oxford Physics Department, Parks Road, Oxford OX1 3PU, UK}
  \author{Z. Tylczynski}
 \address{Institute of Physics, Adam Mickiewicz University,  Umultowska 85, 61-614 Poznan, Poland}
 \date{\today}

\begin{abstract}
We report  nuclear magnetic resonance measurements of the spin-$1/2$ anisotropic triangular lattice antiferromagnet \CsCuCl as a function of temperature and applied magnetic field. 
The observed temperature and magnetic field dependence of the NMR relaxation rate suggests that low energy excitations in the  short-range ordered region stabilized over a wide range  of intermediate fields and temperatures of the phase diagram (sketched in  \mbox{Fig. \ref{Pdiag}(a)}) are gapless or nearly gapless fermionic excitations.
 An upper bound on the size of the gap  of $0.037\, {\rm meV} \approx J/10$ is established.
The magnetization and NMR relaxation rate can be qualitatively described either by a quasi-1D picture of weakly coupled chains, or by mean-field theories of specific 2D spin liquids; however, quantitative differences exist between data and theory in both cases. This comparison indicates that 2D interactions are quantitatively important in describing the low-energy physics. 
\end{abstract}

\maketitle

\section{Introduction}

The emergence of particles with fractional quantum numbers is among the most remarkable phenomena that may arise in strongly correlated electron systems.  In one spatial dimension the exact Bethe ansatz solution of the spin-1/2 Heisenberg antiferromagnetic chain provides a prototype as the solution exhibits a non-magnetic spin liquid ground state with deconfined  spin-1/2 (spinon) excitations \cite{Fadeev81, Haldane83}.  These excitations carry half of the local spin degree of freedom ${\Delta S} =  \pm 1$.    The existence of spinons as fractional elementary excitations has also been well established experimentally in quasi-1D antiferromagnets \cite{tennant95}.   

\begin{figure}[t]
\begin{center}
\centerline{\includegraphics[scale=0.68]{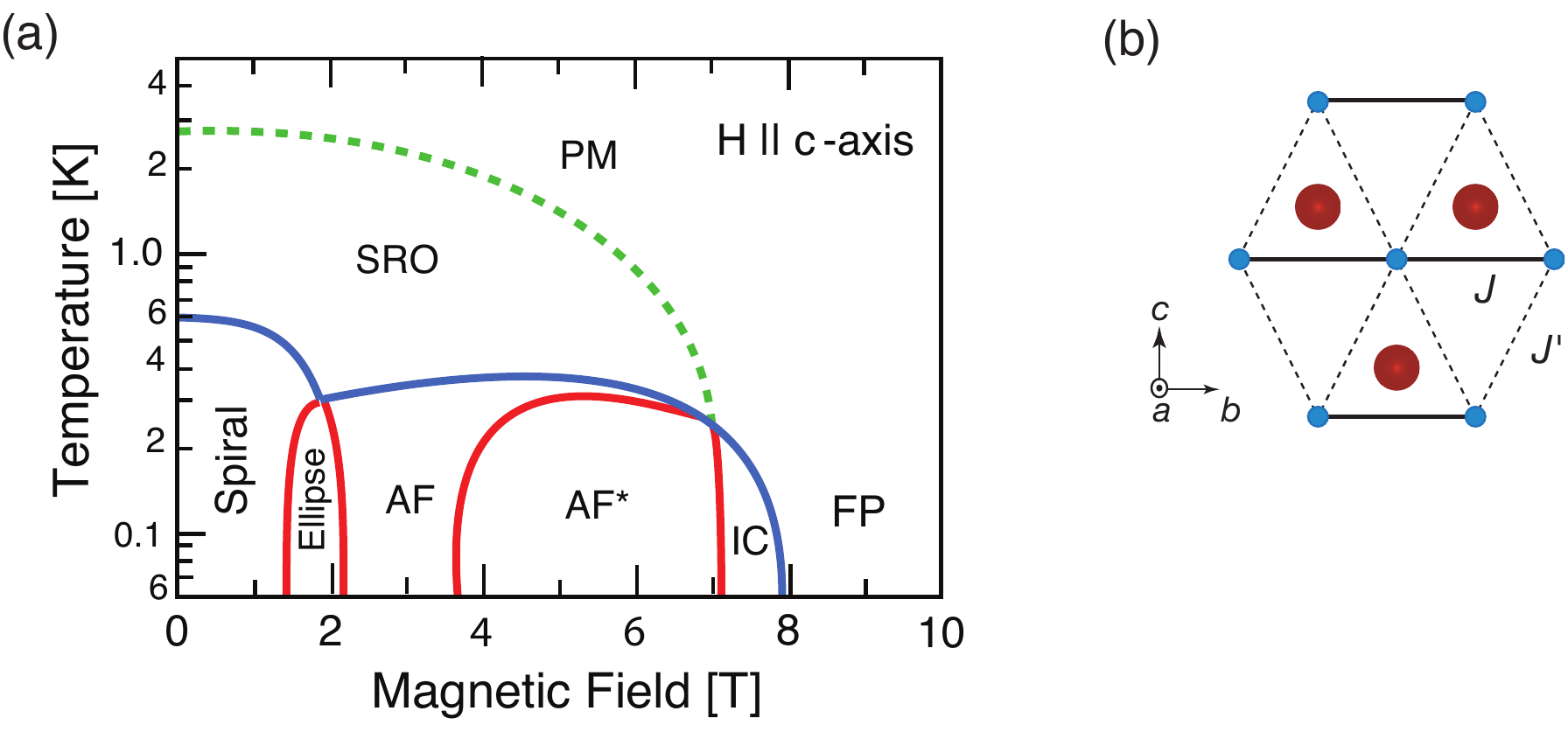}}
\caption{\label{Pdiag}  
\textbf{(a)} Schematic  phase diagram of \CsCuCl for magnetic field applied along the $\hat c$-axis based on results from \cite{Tokiwa:2006, Coldea:PC}. 
The green lines indicate the maximum in the temperature dependence of susceptibility curves and are interpreted as indication of the cross-over from the paramagnetic (PM) to the short range ordered (SRO) region.  First and second order phase transitions are presented by red and blue lines, respectively.   At temperatures below those delineated by blue lines different long range order phases are stabilized. For field above 8 T at  low temperatures fully polarized (FP) state is formed. \textbf{(b)} Triangular magnetic lattice formed by Cu spins (Cu$^{2+}$ ions) displayed as small blue spheres with exchange couplings of $J=0.375$\,meV along solid lines and 
$J'=0.125$\,meV along dashed lines. Our NMR measurements discussed in this paper  were performed on Cs(A) site depicted by red spheres.}
\vspace*{-0.3cm}
\end{center}
\end{figure}

The search for examples of spin fractionalization has turned to dimensions greater than one.  
 Inelastic neutron scattering measurements on the  two dimensional frustrated quantum antiferromagnet (AF) \CsCuCl  with spins on an anisotropic triangular lattice
  have shown dominant continua of excitations as characteristic of \mbox{spin-$1/2$} spinon quasiparticles \cite{Coldea:2001, Coldea:2003} and this has stimulated intense theoretical work to explain these findings \cite{Chung:2001,Zhou:2003,Chung:2003, Sorella:2006, Starykh:2007, Kohno:2007,Kohno:2009}.   \CsCuCl is a hard insulator with orthorhombic space group where
magnetic  spin-1/2 Cu atoms  form a linear chain, with coupling \mbox{$J=0.375$\,meV}, in the $\hat b$-direction. 
Chains are stacked together along the $\hat c$-axis separated by a distance of $b/2$ and with coupling $J'=0.125$\,meV \cite{Coldea:1996,Coldea:2002}.  Thus, the spins form a frustrated anisotropic triangular lattice, as illustrated in \mbox{Fig. \ref{Pdiag}(b)}. Small interplane coupling $J''=0.017$\,meV 
 ($a$-axis) stabilizes long range spiral order below  0.62 K in zero applied field \cite{Coldea:2002}. Schematic  phase diagram of \CsCuCl is shown in  \mbox{Fig. \ref{Pdiag}(a)}. 
A small Dzyaloshinskii-Moriya (DM) interaction $D \sim 5\% J$ (which we neglect in the calculations presented below) 
is present along the interchain links.  At intermediate temperatures ($T \approx  400$\,mK to $2.5$\,K) a short-range ordered (SRO)  region  is stabilized upon application of a magnetic field $(\mathbf{H} = H \hat z)$ of sufficient strength along any of the three crystalline axes \cite{Coldea:2001,Coldea:2003, Tokiwa:2006}.  The magnetic field breaks the full spin-rotational symmetry of an isotropic system, but the subgroup of U(1) rotations in the plane perpendicular to the field remains unbroken in the SRO region.  

Kohno, Starykh, and Balents \cite{Kohno:2007} have shown that many features of  the experimental data in \CsCuCl  can be understood by viewing the anisotropic triangular lattice of spins as a system of weakly-coupled one-dimensional Heisenberg chains.  The work was recently extended to include the effects of an external magnetic field \cite{Kohno:2009} but not yet non-zero temperatures.  As our focus here is on the SRO regime at intermediate temperatures, we also 
compare data to alternative relevant models of spin liquid states previously proposed for the 2D anisotropic triangular lattice, where specific predictions of non-zero temperature properties are available. We consider two  different  spin liquid states, with elementary excitations that obey either bosonic or fermionic statistics,  proposed earlier to describe the spin-1/2 Heisenberg AF on anisotropic triangular lattices. The bosonic Sp(N) large-$N$ mean-field theory supports bosonic spinons that generically have a gap in the excitation spectrum \cite{Chung:2001, Read, SachdevAll}.  By contrast another family of mean-field theories support fermionic spinons with no gap in the excitation spectrum \cite{Chung:2001,Zhou:2003,SBall, Wen:2002}.   In the context of those theories (which assume non-interacting spinons), one can ask which description   (bosons or fermions) yields a better mean-field description.  Those two models have different temperature-dependence of the NMR relaxation rate, which we probe directly in the experiments. 

%
%

Here we measure the local magnetization and the nuclear magnetic resonance (NMR)  relaxation rate to probe the magnetic behavior across the full phase diagram in applied field including the magnetically-ordered phase at low temperatures, the short-range ordered (SRO) region above $T_{N}$,   and the high-temperature paramagnetic region. 
We focus on the properties in the SRO region of the phase diagram where spins are strongly correlated, but not long-range ordered. By considering the temperature and field dependence of the rate we deduce that the low energy excitations in  
 the SRO region are  best characterized as gapless fermionic excitations. Furthermore, we compare our results with previous data on 1D chain materials \cite{Azevedo:1979, Azevedo:1979a} and available theoretical models.  The quasi-1D picture proposed in 
 Ref. \cite{Kohno:2007}  that includes in a consistent way the effects of the strong, but frustrated interchain couplings has not yet been extended to cover the intermediate temperature and field range. As  most of our data was collected in this intermediate 
 temperature and field range, we compare the results with alternative models of spin liquid phases for the anisotropic triangular lattice antiferromagnet with fermionic excitations. Comparison  of the NMR data to mean-field descriptions based upon variational calculations using Gutzwiller-projected wavefunctions implies that  in \CsCuCl in the SRO region (at non-zero temperature and applied field) 2D interactions are important for a quantitative understanding of the low energy properties. 
The rest of the paper is organized as follows. In Sec. II  technical details of the NMR experimental setup are described. In Sec. III we present the results for the temperature and magnetic field dependence of the NMR line and comparison with experiments on 1D materials and with theoretical models. Finally conclusions are summarized in Sec. IV.

\section{Experiment}

We used solution-grown single crystals of \CsE.  The measurements were conducted at the National High Magnetic Field Laboratory (NHMFL)
using a $17$\,T sweepable   magnet. 
We present data on  one of the two magnetically inequivalent Cs sites, labeled Cs(A)  and  believed to be a better probe of  the magnetism of Cu$^{+2}$ ions
due to its stronger hyperfine coupling \cite{Vachon:2006}.
At low temperatures \mbox{($T \lesssim 20$ K)} quadrupolar effects on $^{133}$Cs ($I=7/2$) NMR are masked by the dominant magnetic broadening \cite{Vachon:2006}. The NMR  relaxation rate $(T_1^{-1})$ was measured as described in detail in
 \mbox{Ref. \cite{Vachon:2006}}.  
In essence, the magnetization was saturated by applying  a train of  pulses equally spaced by a time $t  < T_2$ at different frequencies across the magnetically broadened  line. Following the saturation pulse train,  the signal was detected  after a variable delay time    using a standard  spin echo sequence $(\pi/2-\tau-\pi)$.  
The Knight  shift $K \equiv (\omega_{N} - H\gamma)/H\gamma$ is obtained from the frequency of the first spectral moment  $\omega_{N}$ using the  gyromagnetic ratio    \mbox{$^{133}\gamma = 5.5844\, \rm {MHz/T}$}.
The shift also provides a direct measure of the local magnetization $m_{loc} = \mathbf{K} \cdot \mathbf{H}/A_{zz}$,  since they are linearly related via the strength of the transferred hyperfine tensor  $(A_{zz})$  
\cite{Vachon:2008}.  For $ \mathbf{H} || \hat c$, the relevant component of the hyperfine tensor ($(A_{cc})$) for determining $m_{loc}$ equals to $1.23\;  {\rm T}/\mu_{B}$ \cite{Vachon:2008}.  
 The only non-zero off-diagonal element of this symmetric hyperfine tensor is $A_{ac} = \pm 0.185\;  {\rm T}/\mu_{B}$ \cite{Vachon:2008}.

%
\begin{figure}[b]
\begin{center}
\centerline{\includegraphics[scale=0.68]{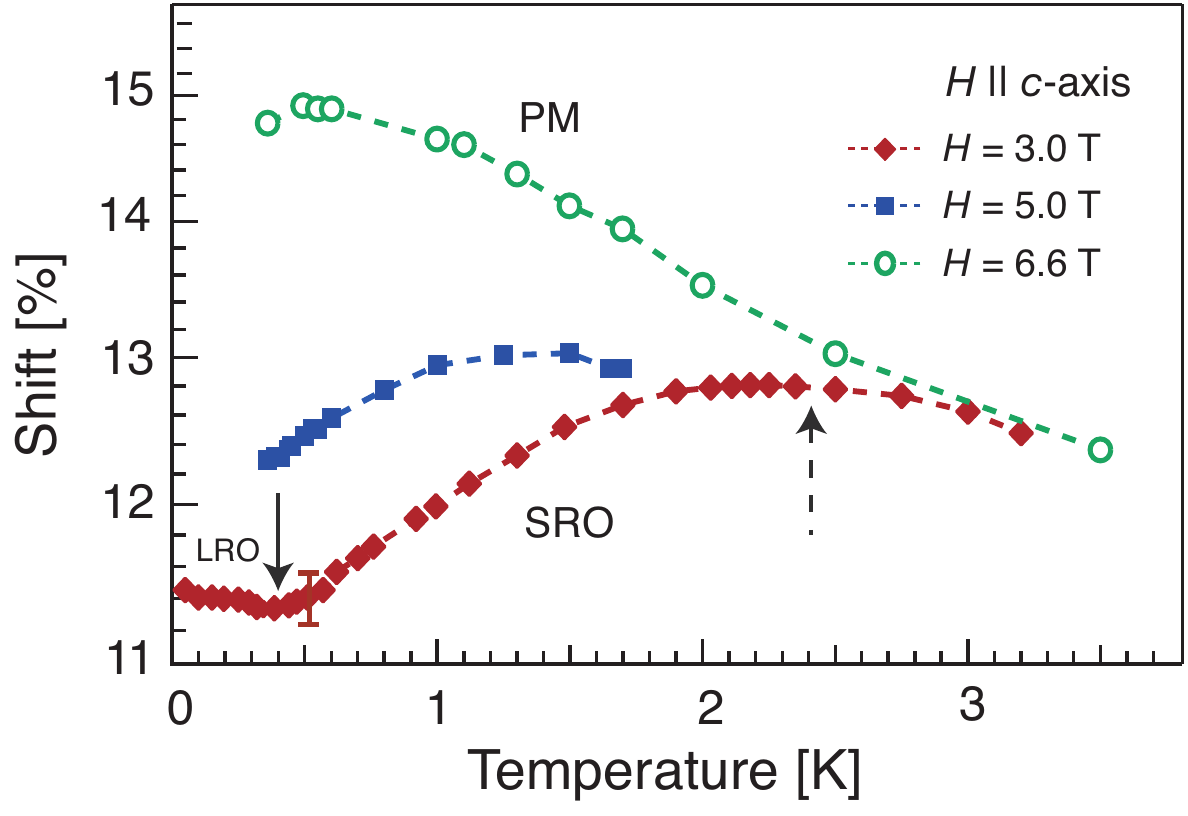}}
\caption{\label{spectra}  
The $T$ dependence of the NMR shift at different values of $\mathbf{H}$ applied along the $\hat c$-axis as denoted.  
A typical error bar is as shown. 
The solid arrow denotes the transition between  LRO (long range order) and SRO (short range order). The dashed arrow denotes cross-over between  SRO (short range order) and PM (paramagnetic) regions.}
\end{center}
\end{figure} 
%

 \section{Results and Discussion}
 
Details of the temperature and applied field dependence of the NMR shift and the rate are discussed in order to delineate different regions  in the phase diagram of  this frustrated magnetic system.  Notably,  we identify the temperature and applied field region where a state characterized by short-range antiferromagnetic correlations is stabilized and we find good agreement with results 
  from bulk magnetization \cite{Tokiwa:2006} and neutron scattering  experiments \cite{Coldea:2001}.  Although this region cannot be distinguished from the paramagnetic one by any change of symmetry, it is appropriate to refer to it as SRO  because spins are strongly correlated but not in a long range ordered state. 
   We now discuss some further ways that our experiment distinguishes between the SRO and paramagnetic regions.  
 
 \subsection{Temperature Dependence}

The temperature dependence  of the shift at different $H$ is illustrated in \mbox{Fig. \ref{spectra}}. 
It exhibits features typical of magnetization of any low dimensional AF with short-range order. 
That is, in the SRO region  the shift increases with increasing $T$, as evident in the \mbox{$H=3$ T} data. This is in contrast to the shift in the paramagnetic (PM) state which decreases with increasing $T$, as apparent in the \mbox{$H=6.6$ T} data. Thus, the characteristic $T$ dependence delineates the boundary to the SRO region. The maximum in the temperature dependence of the shift signals a cross-over from PM to SRO  in the vicinity of $T \approx 2.4 \, {\rm K}$.   This finding is consistent with that from bulk magnetization measurements \cite{Tokiwa:2006}. That is, the low field 
susceptibility, $\chi(T) = M/H$ displays a broad maximum at \mbox{$T \approx 2.8$ K} characteristic of short-range antiferromagnetic correlations \cite{Tokiwa:2006}.  Moreover,   the overall temperature dependence in zero field is well described by high-$T$ series expansion models for the partially frustrated triangular lattice with $J=4.46\,{\rm K}$ and $J'/J=1/3$ \cite{Tokiwa:2006}.

\begin{figure}[t]
\begin{center}
\centerline{\includegraphics[scale=0.54]{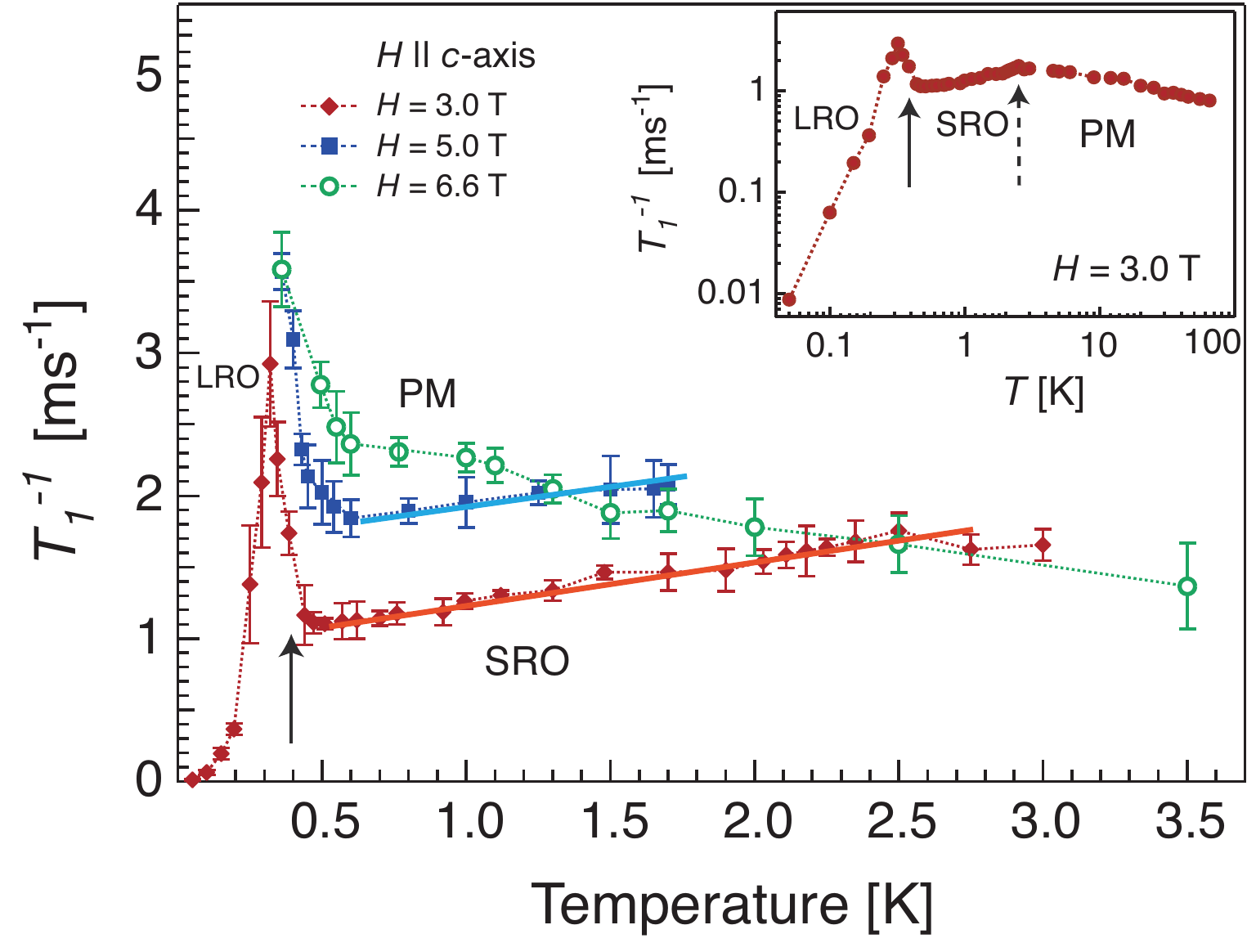}}
\caption{\label{rate_temp} $T_1^{-1}$ as a function of $T$ for different values of   $\mathbf{H}$ applied along the $\hat c$-axis. 
Solid lines denote linear $T$ dependence. Inset: Log-log plot of the $T_1^{-1}$  {\it vs} $T$      at $H=3$\,T. Solid arrow denotes boundary between   LRO  and SRO. 
The dashed arrow denotes cross-over between  regions with different correlations.}
\end{center}
\end{figure}

 The temperature dependence of the NMR relaxation rate also serves to distinguish each 
 region of the phase diagram  as shown in  \mbox{Fig. \ref{rate_temp}}. 
 In the PM region  the rate increases as $T$ is lowered.  The rate reaches its maximum near \mbox{$T \approx 2.4$ K} at $H=3 \, {\rm T}$ as evident in the inset to \mbox{Fig. \ref{rate_temp}}.  
The maximum in $T_1^{-1}$,  like the maximum in the $T$ dependence of the shift, also indicates a cross-over from PM to SRO.   On further lowering   the temperature in the SRO region,  the relaxation rate decreases  linearly in temperature.   
Below  500 mK the rate increases due to an enhancement of fluctuations associated with 
   the transition   to the long-range ordered (LRO) magnetic state.  The exponential decrease   of the rate below \mbox{$T \approx 320$\,mK}  may be due to   the opening of a spin gap in the LRO state (at \mbox{$H = 3$ T}).
%
%
 
  \subsection{Field Dependence}

%
\begin{figure}[b]
\begin{center}
\centerline{\includegraphics[scale=0.64]{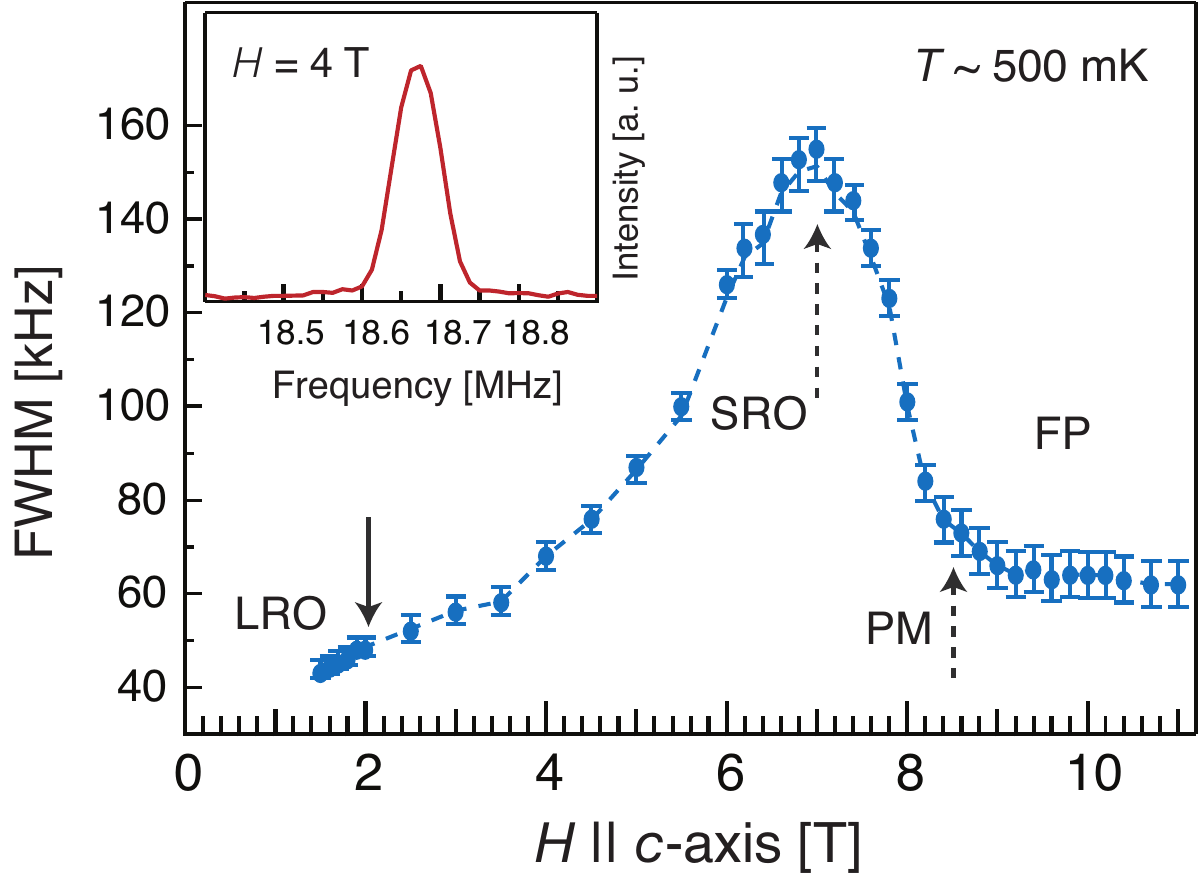}}
\caption{\label{FWHM}  
The applied field   dependence  of the FWHM of the NMR spectra.
Inset:  A representative $^{133}$Cs NMR spectrum  (at \mbox{$T=430$\,mK} and at \mbox{$H = 4$ T})  in the SRO region. 
The solid arrow denotes boundary between  LRO and SRO. 
 The dashed arrows indicate   cross-overs from SRO to PM and then to FP (fully polarized).}
\end{center}
\end{figure} 
%

 The applied field dependence of the full width at half maximum (FWHM)  of  spectra at \mbox{$T\approx 495$\,mK}  is plotted in 
   \mbox{Fig. \ref{FWHM}}. A typical  spectrum, from which FWHM data were extracted,  is shown in the inset to \mbox{Fig. \ref{FWHM}}. 
   At  \mbox{$T\approx 495$\,mK}  for fields below $\approx 7$ T the system is  in the short-range ordered region   as discussed above.  
The FWHM smoothly increases with increasing $H$ up to 
   \mbox{$\approx 7$ T}. 
  As  the FWHM measures the variation of the electron spin  operator projected along the direction of $H$, 
   the increase of the FWHM implies  the increased  static short-range correlation along the direction of $H||c$ in the SRO region. 
  %
   %
\begin{figure}[b]
\begin{center}
\centerline{\includegraphics[scale=0.82]{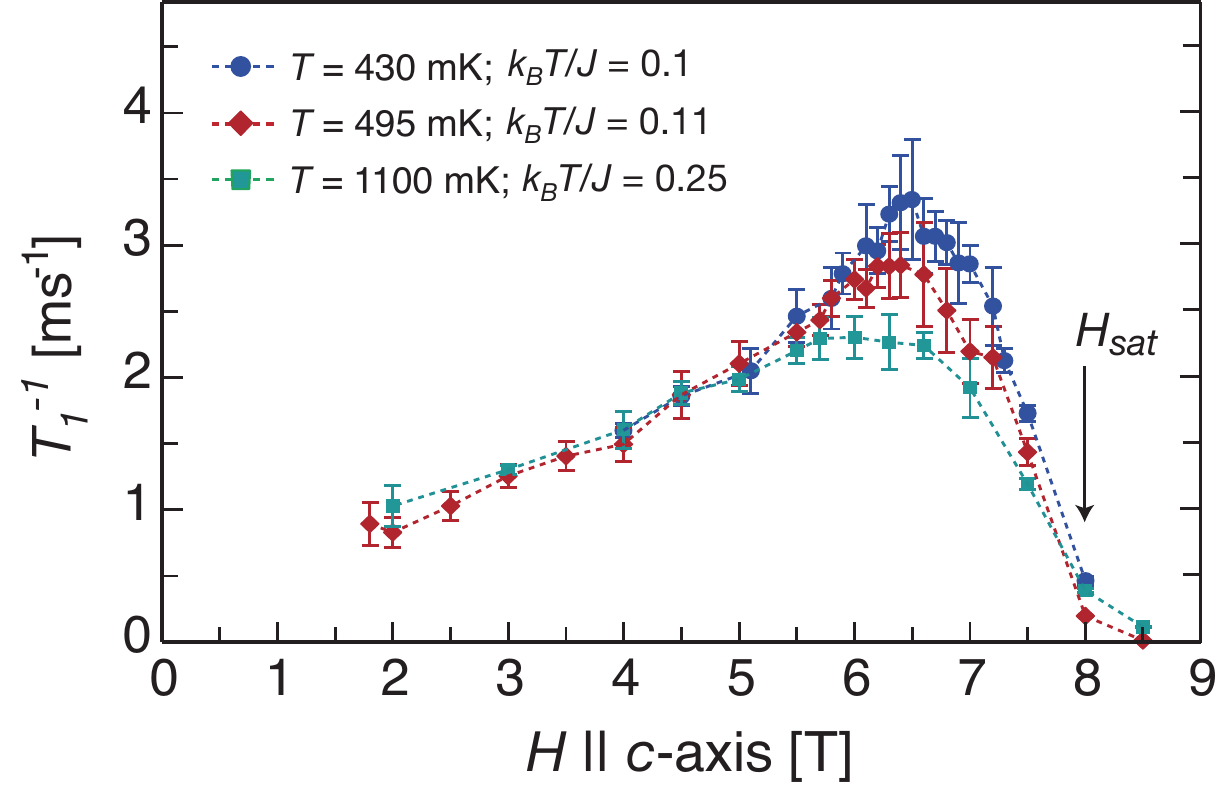}}
\caption{\label{rate_field} The NMR relaxation rate as a function of the 
$H$ applied along the $\hat c$-axis
  at   different temperatures.  }
\end{center}
\end{figure}
%
 %
     We cannot 
   exclude the possibility that some portion of the FWHM increase  with $H$ is due to a DM interaction. However, careful considerations of NMR spectra obtained when $H$ is oriented away from the \mbox{$\hat c$-axis} lead us to conclude that the contribution of DM interaction to the FWHM is  not a dominant one. 
   For $7  \, {\rm T} \lesssim H \lesssim 8.4 \, {\rm T}$, the FWHM decreases with increasing $H$ as expected in the paramagnetic   region, where static short-range correlations are suppressed.   The field at which the FWHM reaches its maximum coincides with that where the maximum of the  differential susceptibility $(d m_{loc}/dH)$ occurs \cite{Tokiwa:2006}.  Thus, the maximum in the FWHM indicates the onset of the cross-over from SRO to PM.    In the fully polarized state the FWHM is constant as all spins are aligned with the field.      
 
 By examining the $T$ and $H$ dependence of the NMR observables we are able to clearly identify  the cross-over from the short-range ordered region.    To gain further insight into the microscopic nature of  this region, we proceed to the analysis of the $H$ dependence of the relaxation rate.   In   \mbox{Fig. \ref{rate_field}}, we plot  the   relaxation rate as a function of applied field.  
  At low fields, as $H$ increases, the rate rises steadily and attains its maximum at a field $H_M$.   The maximum of the rate is smeared out by increasing   
  temperature.  At a given temperature, maxima of both the rate and FWHM occur at nearly equal fields. 

 In the SRO region,  the rate is $T$ independent for $H$ up to  \mbox{$\sim 5$\,T} and, in the limit of $H \rightarrow 0$, it extrapolates to a non-zero value of \mbox{$\sim 0.5$ ms$^{-1}$}. Furthermore,  in the limit of $T \rightarrow 0$ the rate extrapolates to a non-zero value of \mbox{$\sim 1$ ms$^{-1}$} at 3 T  in the SRO region  as evident from the data plotted in \mbox{Fig. \ref{rate_temp}}. 
  The fact that the rate extrapolates to a nonzero value  in the limit of $H \rightarrow 0$ and $T \rightarrow 0$ indicate that the rate is dominated by  gapless excitations. 
Nonetheless, since the SRO region is stabilized  at non-zero temperature we can only place an upper bound on the value of the gap  
$(\Delta)$ at $H=0$.  Specifically, the gap is smaller than the energy scale set by the smallest temperature $(T \simeq 430 \, {\rm mK})$ probed in the SRO region in our experiment,  $k_BT  \simeq 0.037 \, {\rm meV} \approx J/10$. 
Strictly speaking, the evidence of gapless excitations clearly exists only for fields above 1.7 T (for technical reasons this was the lowest field probed in our experiment). 
For one Bohr magneton (Cu$^{2+}$, $S=1/2$) a field of 1 Tesla corresponds to 
$\simeq 0.06 \, {\rm meV} \approx J/6$ which is the same order of magnitude as the energy scale of some of the magnetic couplings in \CsE. It is indeed possible that applying fields larger than 1 T might close a low energy gap due to these weak couplings. However, as we will discuss in the next section, our data agree qualitatively with an observed evolution well-described by gapless 1D fermionic excitations.  Thus, we do not expect additional low energy gaps to open up below 1 T, and even  if that is the case such a gap does not dominate the nature of the quasiparticles over the wide field range of the SRO region. This suggests that the observed behavior of the rate in the SRO is dominated by gapless or nearly-gapless excitations. 

Although the spin dynamics above the ordering temperature is commonly governed by gapless critical fluctuations, the issue  of the existence of a gap in this compound is non trivial  given the fact that the dominant excitations at base temperature occur in a dispersive continuum of excited states 
manifested at intermediate and high  energies,  that remains largely unchanged upon crossing the temperature-driven transitions  from the ordered to the  SRO region 
\cite{Coldea:2001, Coldea:2003}. Moreover, since neutron scattering experiments  were not able to determine a  tight bound on  the gap, it is important to attempt to determine its value with a low energy probe such as NMR as is done in this work \cite{Coldea:2001, Coldea:2003, Chung:2001}.

Having established that the excitations in the SRO region are gapless or nearly gapless, we proceed to examine the statistics of the excitations. The temperature dependence of the NMR rate is particularly sensitive to the statistics of the excitations \cite{Azevedo:1979a, Ehrenfreund:1972}.  
 Assuming that the excitations are weakly interacting, the question is  
 which statistics (bosonic or fermionic) yields a better mean-field description  of the temperature dependence of the rate.  
   The absence of $T$ dependence of the rate for  $H$ up to  \mbox{$\sim 5$\,T} and the 
 fact that the rate extrapolates to a non-zero value  in the limit of $H \rightarrow 0$ and $T \rightarrow 0$  imply that gapless excitations are fermionic \cite{Azevedo:1979}. 
By contrast bosonic excitations  would lead to a much stronger dependence on temperature   \cite{Chung:2003,Azevedo:1979,Ehrenfreund:1972}, in disagreement with the data.  
As discussed in the introduction, the gapless excitations are predicted by fermionic treatments.   A similarly small value of the gap can be obtained from a bosonic treatment if the system is close to a second-order phase transition \cite{Chung:2001};      
however, fermionic spin liquids offer a more natural explanation of the data.

 %
\begin{figure}[t]
\begin{center}
\centerline{\includegraphics[scale=0.65]{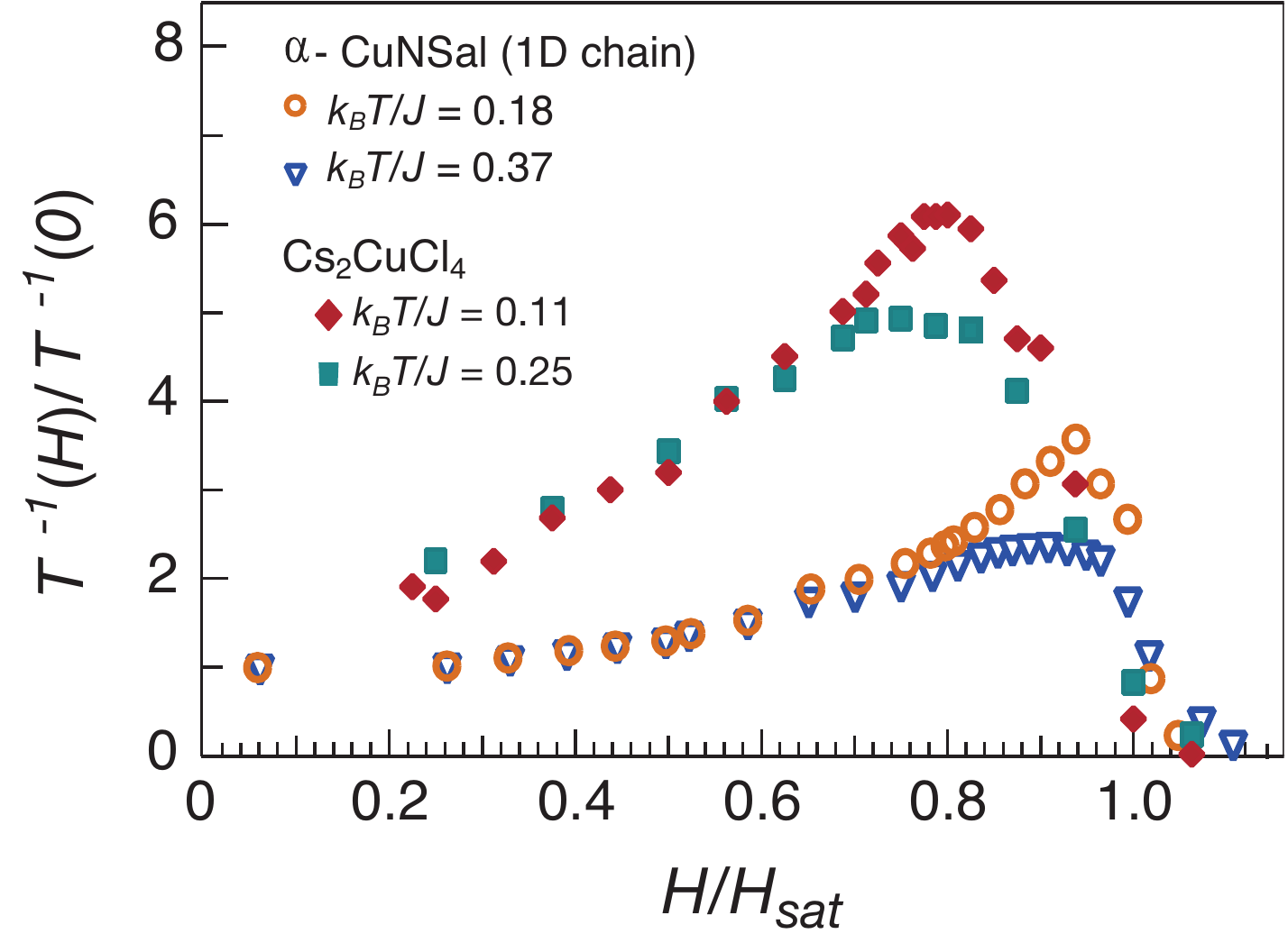}}
\caption{\label{rate1D}  The magnetic field dependence of the NMR relaxation rates measured in  \CsCuCl and in the 1D spin-1/2 AF chain, $\alpha$-CuNSal \cite{Azevedo:1979, Azevedo:1979a}. }
\end{center}
\end{figure}
%

 \subsection{Comparison with the 1D Spin Chain}

The observed field and  temperature   dependence of the relaxation rate resembles that measured in the 1D spin-1/2 AF chain compound  $\alpha$-CuNSal, as illustrated in  
 \mbox{Fig. \ref{rate1D}} \cite{Azevedo:1979, Ehrenfreund:1972}. 
For $H<H_{sat}$, our data agree qualitatively with 
an observed evolution well-described by gapless 1D  fermionic  excitations \cite{Azevedo:1979, Azevedo:1979a}  
suggesting   that the rate in \CsCuCl may be dominated by conventional spinons found in 1D.  On the other hand, there are significant quantitative differences indicating that 
 the strong interchain interactions produce measurable effects. 
  In particular, the peak in the rate is found at $H$ significantly lower than $H_{sat}$, and at low fields the data display a stronger dependence on the field than expected from the model.  
Furthermore, for comparable values of  $k_BT/J$ both the measured and calculated 1D rates \cite{Azevedo:1979,Azevedo:1979a} exhibit  a sharper peak than that evident in the \CsCuCl data.

In \mbox{ Fig. \ref{mag_result}} we show the comparison of the magnetization data with the zero-temperature quasi-1D model proposed by Starykh and Balents in  
 Ref. \cite{Starykh:2007}. 
 This treats the 1D chain exactly and interchain interaction $JÕ$ at mean-field level. Thus, if external field is $H$ then the effective field felt by spins on the 1D chain is slightly smaller, $H_{1d} = H - 4J' \cdot S_{z}$, where $S_{z}$ is magnetization in field $H_{1d}$, $S_z=1/ \pi \arcsin (1/(1- \pi/2+ \pi J/H_{1d}))$, $g=2.3$, $J=0.374$ meV, $J'=0.128$ meV, and $S_z=1/2$ at $H_{1d}=2J$. The disagreement between data and the quasi-1D model appears   at high fields near saturation as the quasi-1D model predicts saturation at a field $2J+2J'$ and not at $2J+2J'+J'^2/2J$, as follows from the spin-wave theory \cite{Tokiwa:2006,Veillette:2005}.  Furthermore effects of the interlayer and DM interaction are not included in the quasi-1D model, nor is rounding due to non-zero temperature $(k_BT/J=0.14)$. For completeness in the section below we also compare the data  with alternative models of spin liquids relevant for the anisotropic triangular-lattice antiferromagnet, where the effects of temperature can be treated explicitly.

 \subsection{2D  Spin Liquid Models}

For the   triangular lattice antiferromagnet a  number of spin liquid states have been theoretically proposed based on the symmetry arguments \cite{Zhou:2003}. Relevant to the anisotropic  triangular lattice and possessing gapless fermionic spinons are two $U(1)$  spin liquids, one with 
commensurate SRO (U1B) and one with incommensurate SRO (U1C). 
A variational study of related states is described in Ref. \cite{Sorella:2006}.  
   Here, we compare the results of static (magnetization) and dynamic (relaxation rate) measurements to predictions based on a combination of Gutzwiller-projected wavefunctions and mean-field theory.  Parameters of the mean-field theory are chosen to minimize the ground state energy of the corresponding Gutzwiller-projected wavefunctions.  For simplicity, the parameters are optimized for the case of zero applied magnetic field and then kept constant over the entire phase diagram.  For  $J/J' \approx 3$,  the optimized  parameters exhibit an enhanced one dimensional character as the ratio of intra-chain to inter-chain hopping amplitudes is  $\lambda/\chi = 7$ and $\lambda/\chi = 8$ for the U1B and U1C states, respectively.  The enhancement of 1D correlations is broadly consistent with the quasi-1D picture of Kohno, Starykh, and Balents \cite{Kohno:2007}.

%
\begin{figure}[b]
\begin{center}
\centerline{\includegraphics[scale=1.15]{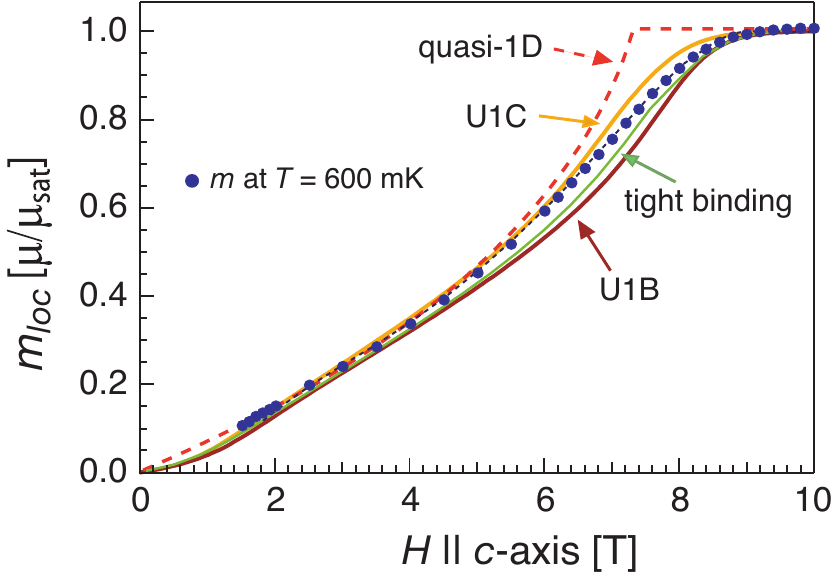}}
\caption{\label{mag_result} Dependence of the magnetization on applied field as calculated  at \mbox{$T=600$ mK} in the two different U(1) spin liquid models (see text) and for the purposes of comparison, a tight-binding model on the anisotropic triangular lattice and a quasi-1D model at zero-temperature described in Ref.\cite{Starykh:2007} for $J=0.374$ meV and $J'=0.128$ meV.  The calculated magnetization is compared to experimental measurements. In the PM regime  $7 \, {\rm T}\, \lesssim H \lesssim 8 \, {\rm T}$, the mean-field theory does not apply and consequently a comparison between measurements and calculations is not meaningful. }
\end{center}
\end{figure}
%

At the mean-field level, the spinons are non-interacting and the magnetic field acts via Zeeman coupling  as a chemical potential of opposite sign for spin-up and down  spinons, changing their relative populations.  We set the bandwidth of the spinon dispersion  so that the critical field required for full polarization of the spins matches that of the experiment in the limit of $T \rightarrow 0$.    A comparison between the calculated magnetization in the different spin liquid states as well as for an anisotropic  tight-binding model,  and the experimentally measured values is shown in \mbox{ Fig. \ref{mag_result}}.  
Similar to the quasi-1D model, both 2D models give a good qualitative description of the data, and the U1C model fits best quantitatively for $H  \lesssim7 \, {\rm T}$. 
 For  $7 \, {\rm T}\, \lesssim H \lesssim 8 \, {\rm T}$, the comparison is not meaningful because there is a cross-over from the SRO to PM regime at \mbox{$\approx$ 7 T} prior to full polarization at \mbox{$\approx$ 8 T}.   The PM state is not captured by mean-field theory nor by the tight-binding model because the bonds have been held constant at their zero-temperature values. 
In addition, the slight disagreement between calculations and data near saturation may also be due to the small inter-layer coupling and DM interaction, both of the order of 
$\sim 5 \% J$,  that  are not included  in the theoretical model and assumed to lead only to a small upward rescaling of the saturation field.

 The relaxation rate for a field oriented along the $z$-direction is given by  \cite{Moriya:1962}:
 \vspace*{0.3cm}
\begin{eqnarray}
 \frac{1}{T_{1z}}&=&\frac{\left(\gamma_{Cs}\mu_B\right)^2}{\hbar^2} \sum_{\beta=x,y}\sum_{i,j} \big(A_{i,x\beta}A^*_{j,x\beta}+A_{i,y\beta}A^*_{j,y\beta}\big)\nonumber\\
&&\times \mathrm{Re} \int^\infty_{0}\mathrm{d}t \exp(i\omega t) \big\langle \big\{\delta S_{i\beta}(t),~ \delta S_{j\beta}(0)\big\}\big\rangle ,
\label{inverset1fromspinspin}
\vspace*{0.3cm}
\end{eqnarray}
where the $A_{i,\alpha \beta} $ denote components of the hyperfine coupling tensor at site $i$ and 
$\delta \mathbf{S}_i=\mathbf{S}_i-\langle\mathbf{S}_i\rangle$.  
Since the $^{133}$Cs(A) ions are located close to the center of a triangle of Cu$^{+2}$ spins \cite{Vachon:2008}, momentum dependent form factors are important and have been included in the calculation.  
With the assumption that  the Cs(A) ions are  located at the center of the triangle,  the form factor  reduces to $\vert A(\mathbf{q}) \vert^2= A^2(3+2\cos(q_x)+2\cos(q_y)+2\cos(q_x-q_y))$ \cite{FormFactors}. 
However, the most important term in the sum in Eq. \ref{inverset1fromspinspin} is due to the transverse ($\beta=x,y$) autocorrelation function ($i = j$) which gives  the following approximate form

\vspace*{0.5cm}
\begin{equation}
 \frac{1}{T_{1z}} \propto \int^\infty_{-\infty}\mathrm{d}\mathcal{E} 
 { {df(\mathcal{E})} \over {d\mathcal{E}} }{ \Big |}_{\mathcal{E}= \mu} \mathcal{N}(\mathcal{E})\mathcal{N}(\mathcal{E}-2\mu), 
\label{approximateinverset1}
\vspace*{1cm}
\end{equation}
where $f(\mathcal{E})$ is the Fermi-Dirac distribution, $\mathcal{N}$ the density of states (DOS), and $\mu$ is the 
Zeeman-shifted chemical potential.   The dominant contribution to the correlation function is proportional to the product of two Fermi functions and is the  result of  scattering two spin-1/2 spinons.  

\begin{figure}[t]
\begin{center}
\centerline{\includegraphics[scale=0.82]{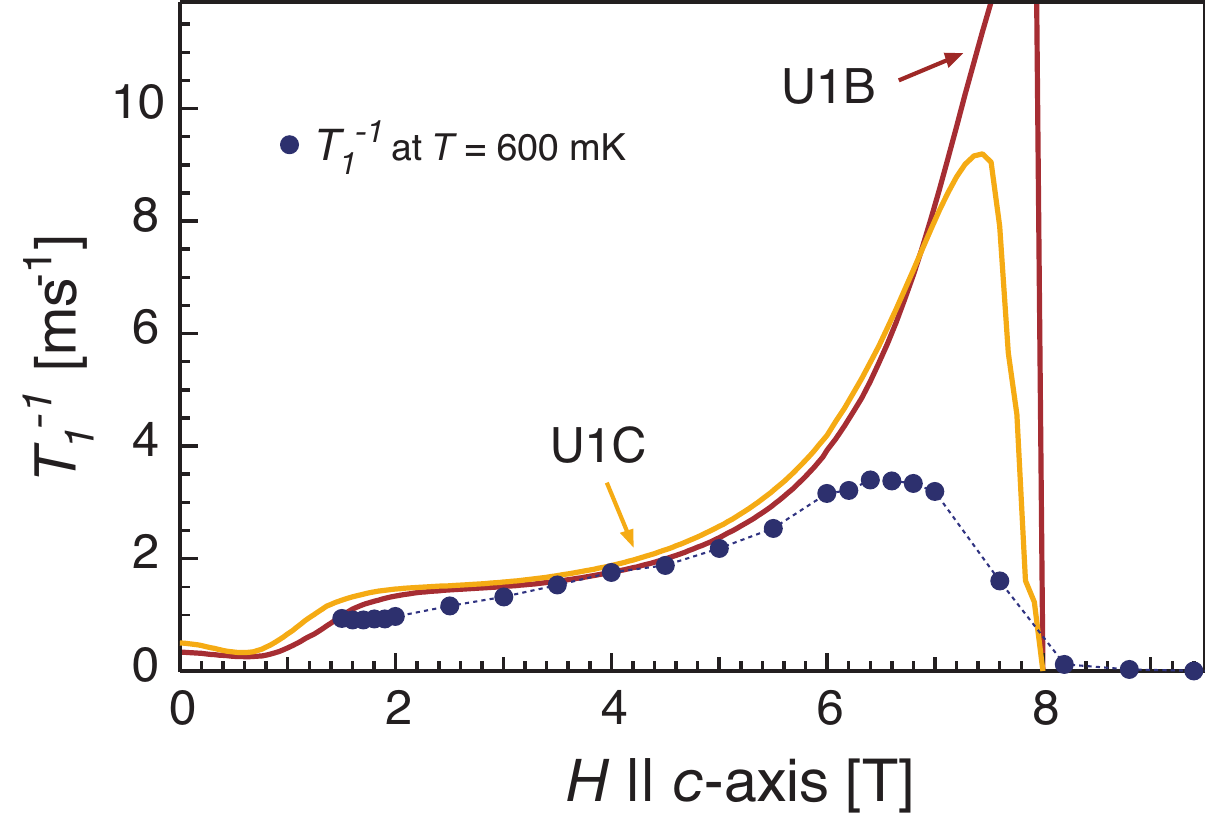}}
\caption{\label{rate_sim} Calculated field evolution of the NMR relaxation rate for the two different U(1) spin liquid models at $T= 600$\,mK using Eq. \ref{inverset1fromspinspin}.
Dark blue symbols are the data.  The dotted curve is a guide to the eye.   }
 \vspace*{-0.5cm}
\end{center}
\end{figure}
%

The   relaxation rate   of the two $U(1)$ SL states is plotted in \mbox{ Fig. \ref{rate_sim}}.  As evident from Eq. \ref{approximateinverset1}, peaks in the relaxation rate correspond to maxima in the DOS.  For the U1C state, the peak occurs prior to saturation at $H \approx 0.9 H_{sat}$, closer to the observed value of $H \approx 0.8 H_{sat}$.  This is because for U1C the DOS is a maximum before saturation is reached, while for U1B the peak is at the $H_{sat}$. 
Unlike  the U1B phase, the DOS for the U1C phase reaches a constant value at the top and bottom of the bands  at saturation.    In contrast, the rate vanishes at low fields in $T \rightarrow 0$ limit in both states because the DOS vanishes at the Fermi surface.  
As calculations are performed at a non-zero temperature of 600 mK, the rate reaches a non-zero value at zero field.  Both models agree with data at intermediate fields but not near saturation where data shows no evidence for a sharp peak/divergence. This may be due to oversimplification in the mean-field approximation for the theoretical models, or the fact that    
the mean-field parameters are optimized for zero field (see above)  so that the high-field behavior can be no more than qualitatively correct.

Important quantitative and qualitative differences between the calculated and the experimentally measured rates remain.
The derivative of  $f(\mathcal{E})$  in \mbox{Eq. \ref{approximateinverset1}} leads to a stronger temperature dependence of the relaxation rate than evidenced experimentally in \mbox{Fig. \ref{rate_temp}}.   In particular,  the rate as calculated in the mean-field approximation vanishes at zero temperature.  By contrast, an extrapolation of the $T_1^{-1}$ SL data plotted in \mbox{Fig. \ref{rate_temp}} down to zero temperature yields  \mbox{$T_1^{-1} \approx 1 \, {\rm ms}^{-1}$} at \mbox{$H=3$ T}. The discrepancy may stem from the fact that the calculated rate, unlike the (one-point) magnetization, is affected by correlations neglected in the mean-field approximation 
\cite{Starykh:2007}.  Gutzwiller-projection is known to change  the power-law exponent of the algebraically decaying spin-spin correlations in SL phases \cite{Arun:2007}.  In 1D, where the rate can be calculated using a Jordan-Wigner transformation to represent spin excitations as spinless fermions, there is only a weak dependence on temperature \cite{Azevedo:1979a}.   Thus, the observed weak dependence of the relaxation rate on temperature may also point towards the dynamics in the SL state  of  \CsCuCl being more like that of 1D spinons traveling along individual chains \cite{Kohno:2007}.
 
 \section{Conclusions}
We have probed the low energy excitations in the  spin-$1/2$ anisotropic triangular  antiferromagnet   \CsE. 
 We have found  that the short-range ordered region stabilized at intermediate temperatures and applied fields is  best characterized by gapless or nearly gapless fermionic spinon excitations, with   experiments yielding an upper bound on the size of the gap value of $0.037 \, {\rm meV} \approx J/10$.  We have compared  the observed field dependence of the magnetization and NMR rate with mean-field theories proposed for anisotropic triangular  lattice antiferromagnet  and found good quantitative agreement for the field-dependence of the NMR relaxation rate at intermediate fields. Our results call for an extension of the self-consistent quasi-1D picture (Ref.\cite{Kohno:2007}) and other theoretical models for spin liquid states in the anisotropic triangular  lattice quantum antiferromagnet to include quantitative predictions for the field and temperature dependence of the NMR  relaxation rate.

\ack{ 
We thank L. Engel, T. Murphey, and E. Palm for help with the experiments. 
The work was supported in part by the NSF (DMR-0547938 and DMR-0605619). A portion of this work was performed at the NHMFL, 
which is supported by NSF Cooperative Agreement No. DMR- 0084173, by the State of Florida, and by the DOE. V. F. M. acknowledges support  by the A. P. Sloan Foundation. }
 


\section*{References}
 \begin{thebibliography}{10}
\vspace{-0.1cm}
\bibitem{Fadeev81} L. D. Fadeev and L. A. Takhtajan,  Phys. Lett. {\bf 85A}, 375 (1981).
\bibitem{Haldane83} F. D. M. Haldane and M. R. Zirnbauer,  Phys. Rev. Lett. {\bf 71}, 4055 (1993).
\bibitem{tennant95} D. A. Tennant, Roger A. Cowley, Stephen E. Nagler, Alexei M. Tsvelik,  Phys. Rev. B {\bf 52}, 13368 (1995).
\bibitem{Coldea:2001} R. Coldea, D. A. Tennant, A. M. Tsvelik, and Z. Tylczynski,  Phys. Rev. Lett. {\bf 86}, 1335 (2001).
\bibitem{Coldea:2003} R. Coldea, D. A. Tennant,  and Z. Tylczynski, Phys. Rev. B  {\bf 68}, 134424 (2003).
\bibitem{Chung:2001} C. H. Chung, J. B. Marston, and R. H. McKenzie,  J. Phys. Condens. Matter {\bf 13}, 5159 (2001). 
\bibitem{Zhou:2003} Y. Zhou and X.-G. Wen, cond-mat/0210662 (2003).
\bibitem{Chung:2003} C.-H. Chung, K. Voelker, and Y. B. Kim, Phys. Rev. B  {\bf 68}, 094412 (2003).
\bibitem{Sorella:2006} S. Yunoki and S. Sorella, Phys. Rev. B {\bf 74}, 014408 (2006).
\bibitem{Starykh:2007} O. A.  Starykh and L. Balents, Phys. Rev. Lett {\bf 98}, 077205 (2007).
\bibitem{Kohno:2007} M. Kohno,  O. A. Starykh, and L. Balents, Nature Physics {\bf 3}, 790 (2007).
\bibitem{Kohno:2009} M. Kohno, Phys. Rev. Lett. {\bf 103}, 197203 (2009).
\bibitem{Coldea:1996} R. Coldea, D.A. Tennant, R.A. Cowley, D.F. McMorrow, B. Dorner, and Z. Tylczynski, J. Phys. Condens. Matter  {\bf 8}, 7473 (1996).
\bibitem{Coldea:2002} R. Coldea, D. A. Tennant, K. Habicht, P. Smeibidl, C. Wolters, and Z. Tylczynski,  Phys. Rev. Lett. {\bf 88}, 137203 (2002).
\bibitem{Tokiwa:2006} Y. Tokiwa,   T. Radu, R. Coldea, H. Wilhelm, Z. Tylczynski, and F. Steglich, Phys. Rev. B {\bf 73}, 134414  (2006).
\bibitem{Coldea:PC}
R. Coldea, {\it private communications}. 
\bibitem{Read} N. Read and S. Sachdev, Phys. Rev. Lett. {\bf 66}, 1773 (1991).  
 \bibitem{SachdevAll} S. Sachdev, Phys. Rev. B {\bf 45}, 12377 (1992); S. Sachdev and N. Read, Int. J. Mod. Phys. B  {\bf 5}, 219 (1991).
\bibitem{SBall} I. Affleck and J. B. Marston, Phys. Rev. B {\bf 37}, 3774 (1988); I. Affleck  {\it et  al.},  Phys. Rev. B   {\bf 38}, 745 (1988);  E. Dagotto, Eduardo Fradkin, and Adriana Moreo,  {\bf {\it ibid.}}  {\bf 38}, 2926 (1988). 
\bibitem{Wen:2002} X.-G Wen,   Phys. Rev. B  {\bf 65}, 165113 (2002).
\bibitem{Azevedo:1979} L.  J. Azevedo, A. Narath, Peter M. Richards, and Z. G. Soos, Phys. Rev. Lett {\bf 43}, 875 (1979).
\bibitem{Azevedo:1979a} L.  J. Azevedo,  A. Narath, Peter M. Richards, and Z. G. Soos, Phys. Rev. B {\bf 21}, 2871 (1980).
\bibitem{Vachon:2006} M.-A. Vachon, W. Kundhikanjana, A. Straub, and V. F. Mitrovi{\'c}, A. P. Reyes, P. Kuhns, R. Coldea, and Z. Tylczynski,    New. J. Phys. {\bf 8}, 222 (2006).
\bibitem{Vachon:2008} M.-A. Vachon, G. Koutroulakis, V. F. Mitrovi{\' c},  A. P. Reyes, P. Kuhns,   R. Coldea, and Z. Tylczynski, J. Phys.: Condens. Matter, {\bf 20}, 295225 (2008).
\bibitem{Ehrenfreund:1972}  E. Ehrenfreund, E. F. Rybaczewski, A. F. Garito,  A. J. Heeger, and P. Pincus, Phys. Rev. B {\bf 7}, 421 (1973).
\bibitem{Veillette:2005} 
M. Y. Veillette, J. T. Chalker, and R. Coldea, Phys. Rev. B {\bf 71}, 
214426, (2005).
\bibitem{Moriya:1962} T. Moriya, Progr. Theoret. Phys., {\bf 16}, 23 (1962).
\bibitem{FormFactors}
 The NMR rate is dominated by the fluctuations that couple via diagonal component of the hyperfine tensor.  Thus,  the off-diagonal components of the hyperfine tensor 
 $A_{ac}$  \& $A_{ca}$, crucial for understanding details of the NMR lineshape (more so in the LRO phases), are less important in determining the NMR rate. 
\bibitem{Arun:2007} A. Paramekanti and J. B. Marston,  J. Phys. Cond. Matt. {\bf 19}, 125215 (2007).
 
\end{thebibliography}
\end{document}